\input{aipcheck}

\documentclass[
    ,final            
  ]
  {aipproc}

\layoutstyle{6x9}

\begin{document}

\title{Leptogenesis in Complex Hybrid Inflation}

\classification{Replace this text with PACS numbers; choose from this list:
\texttt{http://www.aip..org/pacs/index.html}}


\author{Carlos Mart\'inez-Prieto}{
 address={Instituto de F\'isica de la Universidad Michoacana de San Nicol\'as de Hidalgo,
 Morelia Michoac\'an}
}

\begin{abstract}
We study the transference of an initial leptonic charge contained in a complex scalar field (waterfall field) at the end of the inflation
to the leptons of the standard model and then convert this leptonic charge in baryonic charge by sphaleron process.
The proposal is that this is done trough the decay of the complex scalar field particles into the the right-handed neutrino which in turn  decays
 into the left-handed lepton doublet and the Higgs field of the standard model. It must be analyzed in what environment the transference is done.
We propose  that the inflaton (the dominant energy density of the universe) decay into ultrarelativistic fermions before the waterfall field particles decay in the right handed-neutrino, leaving a thermalized bath where the transference of the leptonic asymmetry
can be achieved.
\end{abstract}


\maketitle

\section{Introduction}

It has been shown [1] that we can obtain a $U(1)$ global charge (leptonic) at the end of inflation and that these charge it is associated with the complex scalar waterfall field that ends the inflation. But this leptonic charge must be transferred to the leptons of the standard model. We propose that this it is done
via the disintegration of the watefall field into the right handed-neutrino (right-handed projection of a four component spinor), and later the right-handed neutrino disintegrates into the leptonic doublet and the Higgs doublet of the standard model. The lagrangian that describes these processes is

 \begin{equation}
 \mathcal{L}= h_{1}\bar{\ell}_{L}\Phi N_{R}+ h_{2}\bar{N}_{R}^c N_{R}a, \label{eq:1}
 \end{equation}
where $a$ is the complex waterfall field, $N_R$ is the right-handed neutrino, $\Phi$ is the Higgs Field and $\ell_{L}$ is the left-handed leptonic doublet [2],[3],[4],[5]. \\

But we must analyze in what environment this asymmetry is transferred. When the inflation is finished we have an initial leptonic charge
associated with the waterfall field. Almost instantaneously the scalar fields goes toward the minimum of the potential and then begin to oscillate around it.
At these stage, there is a particle production of the inflaton field, the waterfall field and other fields that interact with them.
If we suppose that the inflaton particles decay into ultrarelativistic fermions with a width or rate of decay much bigger than the widths or rates of decay for the processes of the inflaton with the waterfall field that could suppress the initial leptonic charge, and at the same time this width or rate of decay of the inflaton into a ultrarelativistic fermion is greater than the width or rate of decay of the waterfall field in the right-handed
neutrino,  then the universe can be thermalized before that the watefall field decay, leaving a radiation dominated energy density (the inflaton is the dominant energy density at the end of the inflation and it is converted in  a ultrarelativistic degree of freedom) where the transference of the leptonic charge can be achieved.\\
We study the Boltzmann equations that describe first the transference of the leptonic charge from the initial leptonic charge contained in the Waterfall field
to the right-handed neutrino, then the transference of the leptonic charge of the right-handed neutrino to the lepton doublet of the standard model.
We use only one family of doublets and one right-handed neutrino, but this model can be easily generalized to the three families and three right-handed
neutrinos. Here we must to point out the differences with the standard leptogenesis model [2],[3],[4],[5]. In the standard model of leptogenesis, the lepton asymmetry is obtained trough the decay of right-handed neutrino into the leptonic doublet and Higgs, and the decay of the right-handed neutrino in the lepton antiparticle and Higgs antiparticle (two channels of disintegration for the right-handed neutrino). These processes violate leptonic number and C an CP symmetries and it is out of equilibrium. On the contrary,
in our model these processes  and the others processes involved in the transference, doesn't violate C and CP symmetries, just there is a transference of the asymmetry from  the waterfall field to the leptons of the standard model via the right-handed neutrino. 
This leptonic asymmetry can be converted to a baryonic asymmetry by sphaleron process.
  The asymmetry between baryon and antibaryon can be measured by the quantity
\begin{equation}
B = \frac{\eta_{B}}{s},
\end{equation}
where $\eta_{B}=\eta_{b}-\eta_{\bar{b}}$ that it is the difference between
the baryonic and antibaryonic density, and $s$ is the entropy density.
The astronomical observation give us the numerical constraint [2]
\begin{equation}
6(4)\times 10^{-11} \leq B \leq 1(1.4) \times 10^{-10}.
\end{equation}

\section{The Model of Complex Hybrid Inflation}

The model is given by the potential [1]

\begin{eqnarray}
V(\phi,a) &=& \frac{1}{4\lambda^2} \left( M^2 - \lambda^2 |a|^2
\right)^2 + \left( \frac{m^2}{2} + \frac{g^2}{2} |a|^2 \right) \phi^2
\nonumber \\
&& + \frac{\delta}{4} a^2 \phi^2 + \textrm{c.c.} \label{eq:2},
\end{eqnarray}
where $a$ is the complex Waterfall field, $\phi$ is the inflaton that it is a real field,
g and $\lambda$ are real constants and $\delta$ in general is complex. When $\delta=0$ we obtain the standard hybrid inflation. The delta term violates the
$U(1)$ (leptonic) symmetry. We can eliminate the phase of $\delta$ redefinig $a$, then  we have a CP conserving potential.
The critical points of the potential are: The first
one is a local maximum, and is located at $\phi=|a|=0$, and the
corresponding value of the scalar potential is $V(0,0)
=M^4/(4\lambda^2)$ (false vacuum). The second critical point,
which corresponds to the true vacuum of the system, is a global
minimum and is located at $\phi=0$ and $\lambda |a|/M=1$ (degeneracy).
The $U(1)$ charge density (leptonic) is given by $n_{a}=a_{r}\dot{a_{i}}-a_{i}\dot{a_{r}}$ where the r and i refers to the real and imaginary components of the $a$ field. 

The constant term in the potential~(\ref{eq:2}) is initially
the dominant one. The slow-roll conditions are satisfied and we have the exponential increasing of the scale factor $R=R_{end}exp[H_0(t-t_{end})]$ and the exponential decreasing
of the inflaton field $\phi(t)=\phi_{end}exp[(m^2/3H_0)(t_{end}-t)]$, where $H_0$ is the Hubble parameter during inflation. 
The waterfall field is trapped in a false vacuum $a=0$, but when the inflaton field acquires the value $\phi_-=M/\sqrt{g^2-\delta}$ the imaginary component of the waterfall field presents a tachyonic instability and grows up toward its true vacuum value. When the inflaton field acquires
 the value
 $\phi_+=M/\sqrt{g^2+\delta}$, the real component becomes unstable and tends toward it's true vacuum value, and at the same time inflation is finished. This asymmetric evolution of the components of the waterfall field give us a CP phase necessary to obtain a C and CP violation (are equivalents in this case).
The out of equilibrium is provided by the inflation itself that give us a direction in the arrow of time. These conditions are the condition to obtain
a leptonic charge at the end of inflation. This leptonic charge is of the order of the baryonic charge observed today if the parameters of the model
that satisfied the CMB density perturbations constraints takes values in the ranges given by:
 $\lambda=1$ ,  $0.05<g<0.5$ that correspond to 
$10^{-5}< M/M_{Pl}<10^{-2}$.

\section{Boltzmann equations}
 After the inflation and during the oscillations of the classical scalar fields around the true vacuum, there is a production of particles of the inflaton  and the waterfall fields and fields that interact with them. The parametric resonance production (preheating) is poor [6].\\
A detailed analysis of the evolution of the equation of state ($p=0$ to $p=energy-density/3$) involves a non equilibrium quantum field theory
that is beyond the scope of this paper [7].
The processes that we must consider first are: $a+a \leftrightarrow \phi + \phi$ and $a \leftrightarrow \phi + \phi $.
The masses of the inflaton particles is $Mg$ and of the waterfall particles is $M$.  The rates of decays are $\Gamma_1(a \to \phi+\phi) \simeq \frac{g^4 M}{64 \pi \lambda^2}$, 
$\Gamma_3(a+a \to \phi+\phi) \simeq \frac{g^5M}{32 \pi}$. For $T \ll M$ we have $\Gamma_2(\phi+\phi \to a)=\frac{n_a^{eq}}{n_{\phi}^{eq}}\Gamma_1<\Gamma_1$ and
$\Gamma_4(\phi+\phi \to a+a)=\frac{n_a^{eq}}{n_{\phi}^{eq}}\Gamma_3<\Gamma_3$.

Now, we can suppose that the inflaton field disintegrates into a ultrarelativistic degree of freedom with very small mass compared with the mass of the inflaton
and with a life time short enough that can thermalize the universe before the waterfall field decay into the right-handed neutrino. To do so, we assume that dominant energy density
is the energy density contained in the inflation field.
But we must take into account that the interaction constant in the lagrangian of the interaction between the inflaton field and the ultrarelativistic fermionic particle is small enough to do not induce large loop corrections to the inflaton potential. The interaction lagrangian is 
  $\mathcal{L}_{\phi}= \alpha \bar{\psi}\psi \phi$,
 where $\alpha$ is the interaction constant and $\psi$ is the fermionic field.
The corresponding rate of decay is $\Gamma_{\phi}= \frac{\alpha^2 Mg}{8 \pi}$.

The vev of $a$ add to the lagrangian ~(\ref{eq:1}) a term

$ a \to a+<a> \Rightarrow \mathcal{L}_a= h_{2}\bar{N}^c_{R}N_{R}a+h_{2}\bar{N}^c_{R}N_{R}<a>$,
given $<a>\sim M/\lambda$, then  the mass of the right handed neutrino is $M_{NR}=h_{2}M/\lambda$. We will take $\lambda=1$.

The rate of decay for the process $a \Rightarrow N_R+N_R$ is given approximately by
 $\Gamma_{Da}=\frac{h_2^2M}{8\pi}$.
We need that $\Gamma_{\phi} > \Gamma_{Da}$ to assure that the transference of the leptonic charge to the right handed neutrino will be in a 
thermalized bath of non interacting and dominant energy density of the ultrarelativistic particles. This condition gives to us the constraint
$\alpha^2 g > h_2^2$.  The conditions $\Gamma_1<\Gamma_{Da}$ and $\Gamma_3<\Gamma_{Da}$ can be satisfied by the allowed parameter values , and
the conditions $\Gamma_2 < \Gamma_{\phi}$ and $\Gamma_4<\Gamma_{\phi}$ are satisfied once the condition $\Gamma_{Da}<\Gamma_{\phi}$ is valid.  Then this analysis
show to us that the inflaton decay into the ultrarelativistic fermion as a preferential channel before the waterfall field decay into the right-handed neutrino as it's
preferential channel of decay.\\
Following the analysis of the reference [8] of the decay of the inflaton into ultrarelativistic particles that are the dominant in energy density, we can obtain a reheating temperature and a maximum temperature achieved during reheating. These are given by the formulas
$T_{rh} \simeq 0.1 \sqrt{\Gamma_{\phi} M_{Pl}}$, 
and
 $T_{max} \simeq 0.8 g_*^{-1/4} \left(\frac{M^4 M^3_{Pl}}{4M^6}\right)^{1/2}(\Gamma_{\phi} M_{Pl})^{1/4}$,

where $ \left(\frac{M^4 M^3_{Pl}}{4M^6}\right)$ is the vacuum energy contained in the inflaton field in a Hubble volume at the beginning of the oscillations.
\\
Then at the end of the inflation and at the beginning of the oscillations the inflaton field particles begin to disintegrate into a ultrarelativistic
fermion and when the time $t=\Gamma_{\phi}^{-1}$ they decay rapidly. The processes  due to interactions of the inflaton particles with the waterfall field particles are very suppressed and can be neglected as a first approximation. When the $a$ particle begins to disintegrate rapidly at a time
$t=\Gamma_{Da}^{-1}$, we have a radiation dominate universe, and the dominant processes are $CP$  conserved.
The dominant process are:
   $a \longleftrightarrow N_{R}+N_{R}$, $a^c \longleftrightarrow N^{c}_{R}+N^{c}_{R}$, with width
        $\Gamma_{Da}$.
 $N_{R} \longleftrightarrow  \Phi+\ell_{L}$, $N^{c}_{R} \longleftrightarrow  \Phi^c+\ell^{c}_{L}$,
      with width    $\Gamma_{D}$.
$N_{R} \longleftrightarrow N^{c}_{R}$, with width $\Gamma_{M}$. This term comes from the vev of the waterfall field in the Yukawa coupling of the waterfall field with the right-handed neutrino. This term is treated as an interaction term that converts the right handed neutrino into its antineutrino. Here, we must point out the difference with the standard leptogenesis model. In our model, in the decay of the right handed neutrino,
there is no CP violation, there is just a transference of leptonic number from the right handed neutrino to the leptons of the standard model of particles as we have said before. The mass term for the right-handed neutrino  violates leptonic number and it acts like  a suppression term in the transference
of the leptonic charge as we will see a little later.

Because CPT theorem we have $\Gamma(X \to Y)=\Gamma(X^c \to Y^c)$ for each decay.
 For each species we have  the quantity $Y=n/s$ that  is the number of particles in a comovil volume.

 For the waterfall field we have

 \begin{equation}
 \frac{dY_{a}}{dz_a}=
 -\frac{z_a}{sH(z_a=1)}\left( \frac{Y_{a}}{Y_{a}^{eq}}\gamma(a \to N_{R}+N_{R})
 -\frac{Y_{NR}Y_{NR}}{Y_{NR}^{eq}Y_{NR}^{eq}}\gamma(N_{R}+N_{R} \to a) \right),
\end{equation}
 \begin{equation}
 \frac{dY_{a^c}}{dz_a}=
 -\frac{z_a}{sH(z_a=1)}\left( \frac{Y_{a^c}}{Y_{a^c}^{eq}}\gamma(a^c \to N_{R}^c+N_{R}^c)
 -\frac{Y_{NR^c}Y_{NR^c}}{Y_{NR^c}^{eq}Y_{NR^c}^{eq}}\gamma(N_{R}^c+N_{R}^c \to a^c) \right).
\end{equation}

where $z_a=\frac{m_{a}}{T}$, $H(z_a=1)$ is the Hubble parameter at $T=m_{a}=M$ and $Y_{a}=\frac{n_{a}}{s}$, $Y_{a^c}=\frac{n_{a^c}}{s}$, $Y_{NR}=\frac{n_{NR}}{s}$,$Y_{NR^c}=\frac{n_{NR^c}}{s}$ .

Here
\begin{equation}
\gamma(aX \to Y)=\int d\pi_a d\pi_X d\pi_Y (2\pi)^4 \delta^{4}(p_a+p_X-p_Y) f_{a}^{eq}f_{x}^{eq}\vert M(aX \to Y)\vert^2,
\end{equation}
where $f_a^{eq}=e^{-E_a/T}$ [3].
\\
If $CP$ is conserved and because the energy conservation during the process, we have $\gamma(aX \to Y) = \gamma(Y \to aX)$. The gamma is given in terms of the width by the equations
\begin{equation}
 \gamma(a \to Y)=n_a^{eq}\langle \Gamma(a \to Y)\rangle,
\end{equation}
\begin{equation}
 \langle \Gamma(a \to Y)\rangle=\frac{K_1(m_a/T)}{K_2(m_a/T)}\Gamma(a \to Y).
\end{equation}
For two body scattering we have
\begin{equation}
 \gamma(ab \to Y)=n_a^{eq} n_b^{eq} \langle \sigma(ab \to Y)\vert v \vert \rangle,
\end{equation}

where $K_{1}(z)$ and $K_{2}(z)$ are the modified Bessel functions and $\Gamma$ is the usual decay  width at zero temperature in the rest system of the decaying particle. 

The equilibrium or out of equilibrium for a kind of particle is given by the next rule: Defining $K=\Gamma/H\vert_{T=mass}$, we can say that the particle is out of equilibrium if $K\leq1$, and it is in equilibrium if $ K>1$.
The rates of decay are: 

 $\Gamma_{Da}=\frac{h_{2}^{2}M}{8\pi}$, $\Gamma_{D}=\frac{\tilde{m}_{1}M^2_{NR}}{8\pi v^2}$, $\Gamma_{M}=M_{NR}$.

 where $8 \times 10^{-12} GeV < \tilde{m}_{1}<5 \times 10^{-11} GeV$,  $v=174$ GeV.

 Defining the variables $z_{a}=M/T$ and $z=M_{NR}/T$ and taking $h_2=6.25\times10^{-4}$ and $M \sim 10^{16}GeV$, we have
$K_{a}=(\Gamma_{Da}/H(z_{a}=1))\sim 10^{-6}$, it is out of equilibrium. 

For the decay of $NR$ we have $K_{D}= \Gamma_{D}/H(z=1) \sim 1.3 \times 10^2$, $K_M\sim 4\times 10^3$ it is in equilibrium.
\\
\subsection{Solution to Boltzmann Equations}
 Taking into account the CP conservation and the energy conservation and writting the $\gamma$ in terms of the width, the density functions and the time dilation for the width
($K_1(z)/K_2(z)$) and considering that the right handed neutrino goes toward equilibrium we have for the waterfall field
\begin{equation}
 \frac{dY_{a}}{dz_a}=-\frac{\Gamma_{Da}}{H(z_a=1)}\frac{z_aK_1(z_a)}{K_2(z_a)}\left(Y_a-Y^{eq}_a\right).
\end{equation}
Making the same considerations for the right-handed neutrino and assuming equilibrium for the product of decays we obtain

\begin{eqnarray}
\frac{dY_{NR}}{dz}=&-&\frac{\Gamma_{Da}}{H(z=1)}\frac{zK_1(z)}{K_2(z)}\left( Y_a^{eq}-Y_{a}\right) \nonumber \\
&-&\frac{\Gamma_D}{H(z=1)}\frac{zK_1(z)}{K_2(z)}\left(Y_{NR}-Y^{eq}_{NR}\right) \nonumber \\
&-&\frac{\Gamma_M}{H(z=1)}\frac{zK_1(z)}{K_2(z)}\left(Y_{NR}-Y_{NR^c}\right).
\end{eqnarray}

Finally with the same considerations about the kinetic equilibrium we obtain for the doublet
\begin{equation}
\frac{dY_{l}}{dz}=\frac{\Gamma_{D}}{H(z=1)}\frac{zK_1(z)}{K_2(z)}\left(Y_{NR}-Y_{NR}^{eq}\right).
\end{equation}
The equations for the antiparticles are exactly the same because by CPT the widths are the same.
 By CPT,  $Y^{eq}_p=Y^{eq}_{p^c}$.
Defining   $\Delta_{b}=Y_{b}-Y_{b^c}$ and
taking $K_{1}(z)/K_{2}(z)=z/(2+z)$ for the Bessel  functions,
we have two regimes:
$z<1$ that implies $ K_1(z)/K_2(z)\sim z/2$ and $z>1$ that implies $K_1(z)/K_2(z) \sim 1$. 
We expect that at $T=T_{rh}$ the waterfall field $a$ decay into the right handed neutrino, that is when  ($z_a=M_a/T=M/T>1$). And the decay
of the right handed neutrino it is achieved at $z=M_{NR}/T\ge 1$. The relation between the two variables is $z_a=M/T=(M/M_{NR})(M_{NR}/T)=hz$

For $z>1$ we have the equations

\begin{equation}
 \frac{d\Delta_{a}}{dz_{a}}=-K_{a}z_a\Delta_{a}.
\end{equation}
It's solution with an initial condition $\Delta_a(0)=\Delta_a^0$ is
\begin{equation}
\Delta_a(z_a)=\Delta_a^0 e^{-K_az_a^2/2}.
\end{equation}
In terms of $z$ the solution becomes

\begin{equation}
 \Delta_{a}(z)=\Delta_a^0 e^{-K_{a}h^2 z^2/2}.
\end{equation}

For $\Delta_{NR}$  we have

\begin{equation}
 \frac{d\Delta_{NR}}{dz}= z(h^2K_{a}\Delta_{a}-K_{DM}\Delta_{NR}),
\end{equation}

where $K_{DM}=(\Gamma_{D}+2\Gamma{M})/H(z=1)$. The solution with initial condition $\Delta_{NR}(0)=0$ is

\begin{equation}
 \Delta_{NR}(z)=\frac{h^2K_{a}\Delta^0_a}{K_{DM}-h^2K_a}\left(Exp(-h^2K_az^2/2)-Exp(-K_{DM}z^2/2)\right).
\end{equation}

For $\Delta_{l}$ we have 

\begin{equation}
\frac{d\Delta_{l}}{dz}=z K_{D}\Delta_{NR}.
\end{equation}

The solution with initial condition $\Delta_l(0)=0$ is

\begin{equation}
 \Delta_l(z)=\frac{h^2K_{D}K_a\Delta_a^0}{K_{DM}-K_ah^2}\left(\frac{e^{-K_{DM}z^2/2}}{K_{DM}}-\frac{e^{-K_ah^2z^2/2}}{K_ah^2}\right)+\frac{K_D\Delta_a^0}{K_{DM}}.
\end{equation}
The final leptonic asymmetry  is suppressed with respect to it's initial value in the scalar sector, and  is given by
\begin{equation}
 \Delta_l(z=\infty)= \frac{K_D\Delta_a^0}{K_{DM}}=\frac{1}{1+2\frac{\Gamma_{M}}{\Gamma_{D}}} \Delta_a^0.\label{eq:3}
\end{equation}
The suppression factor can be given in terms of the quotient $\Gamma_M/\Gamma_D$.
We must consider that
\begin{equation}
 \Delta_a^0= 0.56 \left( \frac{M}{M_{Pl}}\right)^2 xe^{-x^2} 
\end{equation}
where $x^2 \simeq 2.93 \times 10^{-4}/g^2$ [1].

\begin{figure}
               \includegraphics[width=8cm]{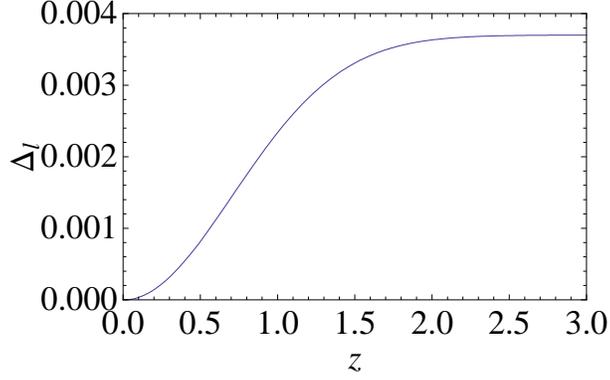}   
               \caption{Leptonic charge in the doublet versus z.}
\end{figure}

\section{Conclusions}
It has been  developed a mechanism of transference of a leptonic charge contained in the waterfall field of the complex hybrid inflation 
to the leptons of the standard model via the right-handed neutrino. To assure that the transference takes place in a thermalized
environment, the proposal is that the inflaton decay in a ultrarelativistic fermion leaving a radiation dominated environment where the transference
of the leptonic charge takes place. The study of the Boltzmann equation of the transference of the leptonic charge, show to us that we have a
suppressed final leptonic charge, and these suppression factor depends of the quotient $\Gamma_M / \Gamma_D$. 
If we consider that $B \sim 0.4 \vert \Delta_l(z=\infty)\vert \sim 10^{-10}$ then the parameter values that satisfy all the constrains are:
$\lambda=1$, $g \sim 2.42 \times 10^{-2}$, $\alpha \sim 10^{-2}$, $M \sim 10^{16}GeV$ and $h_2 \sim 6.25 \times 10^{-4}$. These values yields to
a reheating temperature $T_r \sim 10^{13}GeV$ and a right-handed neutrino mass $M_{NR} \sim 6.25 \times 10^{12}GeV$.

\begin{theacknowledgments}
We thank David Delepine for useful discussions in central points of this work, to Luis Ure\~na and Ulises Nucamendi for useful discussions, and the
  organizers of the III International Meeting on Gravitation and Cosmology held in Morelia Michoac\'an last May for the opportunity to present these
  results. This work was supported by CONACYT Postdoctoral program.
\end{theacknowledgments}

\bibliographystyle{aipproc}

\end{document}